\documentclass{moriond}

\def\Journal#1#2#3#4{{#1} {\bf #2}, #3 (#4)}

\def\be{\begin{equation}}
\def\ee{\end{equation}}
\def\bea{\begin{eqnarray}}
\def\eea{\end{eqnarray}}

\usepackage {siunitx}
\usepackage {braket}
\usepackage {amsmath}
\newcommand\qbounce{\textit{q}B{\sc{ounce}} }
\newcommand\lb{\left(}
\newcommand\rb{\right)}

\newcommand\ibc[2]{$\ket{#1}\rightarrow\ket{#2}$}
\newcommand\ebc[2]{\ket{#1}\rightarrow\ket{#2}}

\newcommand{\Photo}{\includegraphics[height=35mm]{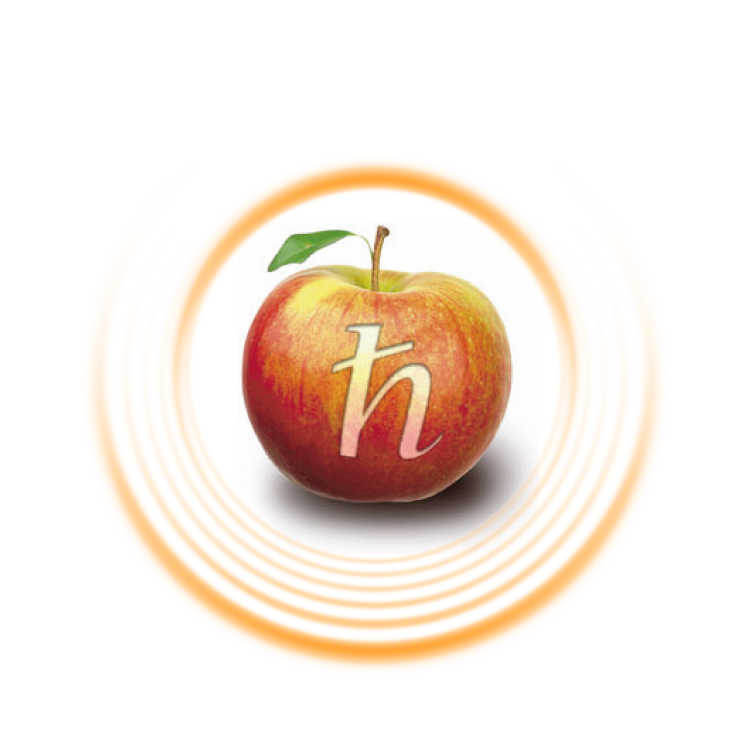}}

\begin{document}
\vspace*{4cm}
\title{\qbounce: Systematic shifts of transition frequencies of gravitational states of ultra-cold neutrons using Ramsey gravity resonance spectroscopy}
\author{ Jakob Micko\,\footnote[1]{Institut Laue-Langevin (ILL), Grenoble, France}$^,$\,\footnote[2]{Atominstitut, TU Wien, Austria}, J. Bosina\,\footnotemark[2], S. S. Cranganore\,\footnotemark[2], T. Jenke\,\footnotemark[1], M. Pitschmann\,\footnotemark[2], S. Roccia\,\footnotemark[1], R.I.P. Sedmik\,\footnotemark[2], H. Abele\,\footnotemark[2]}

\address{ILL, Grenoble, ATI TU Wien}

\maketitle\abstracts{
\qbounce is using quantum states of ultra-cold neutrons in the gravitational field of the Earth to investigate gravitation in the micrometre range. We present current measurements taken in 2021 at the Institut Laue-Langevin (ILL) to determine energy differences of these states by mechanically induced transitions. This allows a determination of the local acceleration $g$ using a quantum measurement. The data presented here results in $g=\SI{9.812+-0.0018}{\metre\per\second^2}$. The classical local value at the experiment is $g_c=\SI{9.8049}{\m\per\second^2}$. We present an analysis of systematic effects that induces shifts of the transition frequency of order \SI{100}{\milli\hertz}. The inferred value for $g$ at the experiment shows a systematic shift of $\delta g\approx3.9\sigma$.}

\section{Introduction}
Neutrons are excellent probes to test gravity at short distances – electrically neutral and only hardly polarizable. Very slow, so-called ultra-cold neutrons (UCN) form quantum states bound by the gravity potential of the Earth. This allows combining gravity experiments with powerful resonance spectroscopy techniques, as well as tests of the interplay between gravity and quantum mechanics. Recently, the \qbounce collaboration has finished commissioning its Ramsey spectrometer at the ultra-cold and very cold neutron facility PF2 at the Institut Laue-Langevin. Here, we present first precision measurements, its result and an analysis of systematics effects.

\begin{figure}[h]
\includegraphics[width=454 pt]{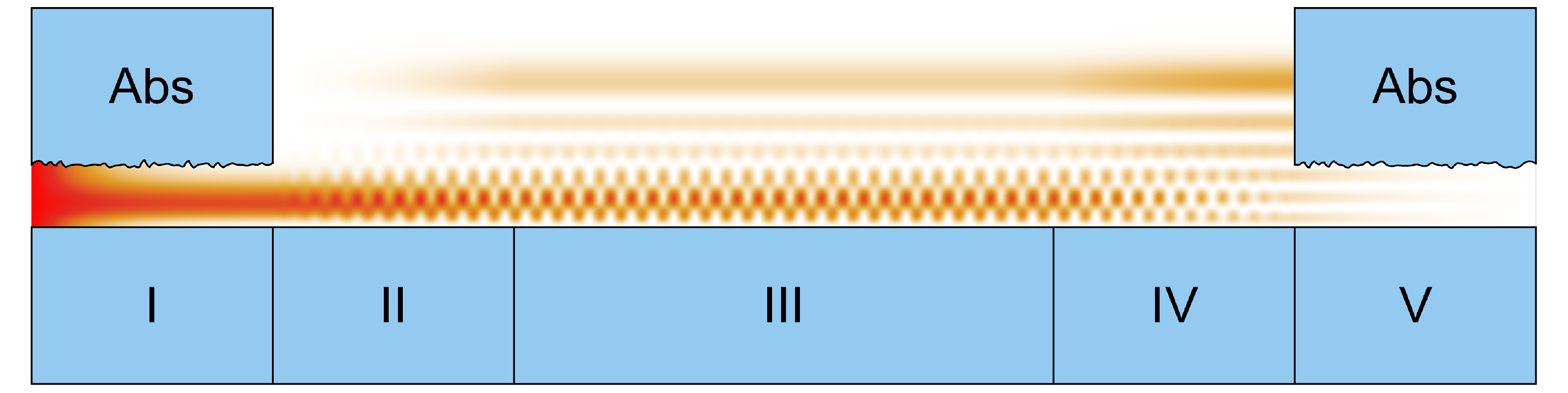}
\caption{Schematic view of the Ramsey spectrometer. UCNs enter from the left and are detected on the right. The probability density function of the neutron in the setup is shown in false colour.}\label{fig:qbschematic}
\end{figure}

A schematic overview of the experiment can be seen in Fig. \ref{fig:qbschematic}. The neutrons have a horizontal velocity of $v_x\approx\SI{8}{\metre\per\second}$ and a vertical velocity of $v_z\approx\SI{0}{\meter\per\second}$. They are trapped by the gravity potential and form bound states \cite{nesvizhevsky2002}. 
In section I neutrons enter the system on top of a flat surface. They form bound states in the gravity potential with energies in the order of \SI{}{\pico\eV} for details see contribution by J. Bosina in this volume. The Fermi pseudo potential of the mirror is $\approx\SI{100}{\nano\eV}$ and the ground state extends $\approx\SI{15}{\micro\meter}$ above the mirror, with higher states extending further up. On top of section I a rough glass plate (absorber) is placed at a distance of $l\approx\SI{25}{\micro\meter}$ which causes high states to be scattered. This scattering happens on the random surface structure and is non specular. The scattered neutrons have a random velocity in the x-y plane and do not enter the detector at the end of the experiment. The scattering probability of low state neutrons is lower and after section I the wavefunction is in a superposition of the first three states with $(P(1), P(2), P(3))^T\approx (0.70, 0.25, 0.05)^T$. A detailed discussion can be found in \cite{Chizhova2014}.
In section II the mirror oscillates with a specific frequency, displacement and phase where the surface position is given by $z_{II}(t)=a_{II}\sin\lb\omega_{II} t+\phi_{II}\rb$. In Fig. \ref{fig:qbschematic} the transition \ibc{1}{6} is depicted. This is the transition (mainly) investigated in 2021. If the lower and the excited state have an equal probability at the end of section II the applied excitation is called a $\frac{\pi}{2}$-flip. 
Section III is stationary and the neutron propagates without perturbation.
In section IV the mirror is oscillating with $z_{IV}(t)=a_{IV}\sin\lb\omega_{IV} t+\phi_{IV}\rb$, where $\omega_{IV}$ and $a_{IV}$ are tuned to be the same frequency and displacement as in section II. If sections II and IV are in phase, such that $\phi_{IV}=\phi_{II}$ the excitation to the higher state is completed. If $\phi_{IV}=\phi_{II}+\pi$ the neutron returns to the ground state. 
In section V the same selection as in section I is performed.
This enables us to perform Ramsey type spectroscopy using Gravity states a technique called Gravity Resonance Spectroscopy (GRS). From data on the measured transition probability the local acceleration of the neutron can be determined.

\section{Theory}
The evolution of the system can be described by the Schrödinger equation:
\begin{equation}
i\hbar\partial_t \psi=\hat{H} \psi
\end{equation}
For a linear potential the solutions can be given in terms of shifted Airy functions. With linearised Newtonian gravity the solutions only depend on the local acceleration $g$ and the mass of the neutron $m_n$. 

\qbounce involves infinitely many states which makes precise numerical calculations necessary. The dominant contribution for this dataset is \ibc{1}{6} (and \ibc{2}{7}). A simplified two state system, the lower energy one is called $\ket{p}$ and the higher one $\ket{q}$ can be solved analytically. The detailed derivation of the transition probability from one state to the other can be found for example in \cite{rechberger2018}. For this as an illustrating example the probability to remain in the lower state is given by
\begin{equation}\label{eq:ramsey}
\begin{split}
&P(\ebc{p}{p})=\left|a_{II}a_{IV}-b_{II}b_{IV}e^{i\lb\delta_{IV}\lb T+\tau\rb-\delta_{II}\tau-\lb\chi_{II}+\chi_{IV}\rb+\phi_{IV}-\phi_{II}\rb}\right|^2\,,\\
&\Omega_{R,i}=\sqrt{\delta_i^2+\lb a\omega_i|V_{pq}|\rb^2},\;\chi_i=\arctan\lb \frac{\delta_i}{\Omega_{R,i}}\tan\lb\tau\frac{\Omega_{R,i}}{2}\rb\rb\,,\delta_i=\omega_i-\omega_{qp}=2\pi\lb\nu_i-\nu_0\rb\\
&a_i=\sqrt{\cos^2\lb\tau\frac{\Omega_{r,i}}{2}\rb+\frac{\delta_i}{\Omega_{R,i}}\sin^2\lb\tau\frac{\Omega_{r,i}}{2}\rb},\; b_i=\frac{a\omega_i|V_{qp}|}{\Omega_{R,i}}\sin\lb\tau\frac{\Omega_{r,i}}{2}\rb\,,
\end{split}
\end{equation}
where $a\omega_i$, $\omega_i=2\pi\nu_i$, $\nu_i=\frac{\omega_i}{2\pi}$ are the oscillation velocity, the angular frequency and the frequency for the i-th section respectively, $V_{qp}$ is the overlap integral for the perturbation, $\nu_0=\frac{\omega_{qp}}{2\pi}$ is the transition frequency, $\tau$ is the interaction time in sections II and IV and $T$ is the free propagation time in section III. Simplifying to the case that both sections oscillate with the same strength $a\omega_{II}=a\omega_{IV}=a\omega$ and frequency $\nu_{II}=\nu_{IV}=\nu$ and $\nu-\nu_0$ is small, gives
\begin{equation}\label{eq:ramseyprob}
P(\ebc{p}{p})=1-\frac{\sin^2\lb a\omega |V_{pq}|\rb}{2}\lb1+\cos\lb2\pi (\nu-\nu_0)\lb T+\frac{\tan(\frac{a\omega |V_{pq}|}{2}\tau)}{\frac{a\omega |V_{pq}|}{2}}\rb+\Delta\phi\rb\rb
\end{equation}
where $\Delta\phi=\phi_{IV}-\phi_{II}$. In the case that section II and IV are not oscillating with the same vibration strength roughly at resonance $\nu=\nu_0$ eq. (\ref{eq:ramsey}) for two different vibration strengths simplifies to
\begin{equation}\label{eq:transprob}
\begin{split}
P=\left|\cos\lb\tau \frac{a\omega_{II}V_{pq}}{2}\rb\right.&\left.\cos\lb\tau\frac{a\omega_{IV}V_{pq}}{2}\rb-\sin\lb\tau\frac{a\omega_{II}V_{pq}}{2}\rb\sin\lb\tau\frac{a\omega_{IV}V_{pq}}{2}\rb e^{i\Delta\phi}\right|^2\;,\\
\Rightarrow
&P(\ebc{p}{p})=\cos^2\lb\frac{\tau V_{pq}}{2}\lb a\omega_{II} \pm a\omega_{IV}\rb\rb\;,
\end{split}
\end{equation}
with the upper sign for $\Delta\phi=0$ and the lower sign for $\Delta\phi=\pi$.

\section{Experiment}

\begin{figure}
\includegraphics[width=454 pt]{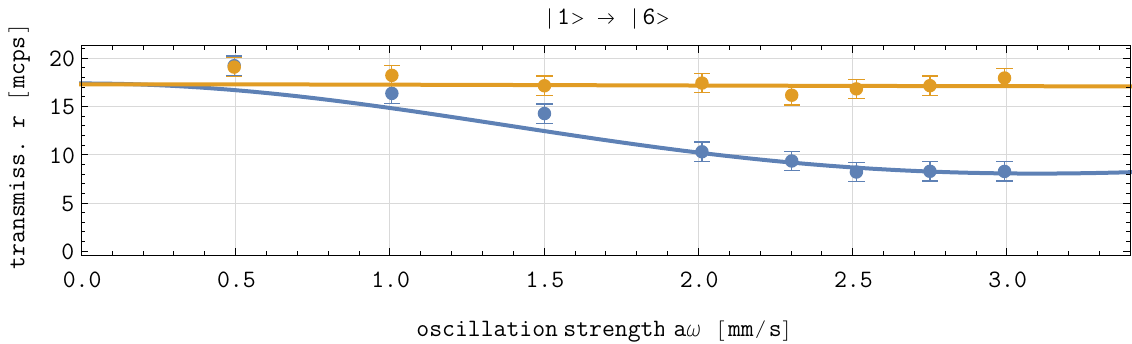}
\caption{Determination of the optimal $\pi/2$-flip by increasing the oscillation strength $a\omega$. $\Delta\phi=0$ in blue and $\Delta\phi=\pi$ in orange. The solid lines show the theory function fitted to the measurements.}\label{fig:aw16sweep}
\end{figure}

During the second reactor cycle at PF2 at the ILL in 2021 the transition \ibc{1}{6} was studied in detail. As an initial guess the transition frequency was calculated using the classical value of $g\approx\SI{9.805}{\meter\per\second^2}$ gives an energy of $E_1= \SI{1.407}{\pico\eV}$ $E_6= \SI{5.428}{\pico\eV}$ which results in the transition frequency $\nu_{16}=\SI{972.345}{\hertz}$. The oscillating sections were set to the same frequency and amplitude with a phase difference of $\Delta\phi=0$ in (\ref{eq:transprob}), all of which was measured with an interferometer. The oscillation strength was increased until a minimum in the transmission rate through the setup was observed. As a consistency check the same frequencies were also measured using $\Delta\phi=\pi$. These measurements can be seen in Fig. \ref{fig:aw16sweep} with the rate given in counts per 1000 seconds [mcps]. 

\begin{figure}[h]
\includegraphics[width=454 pt]{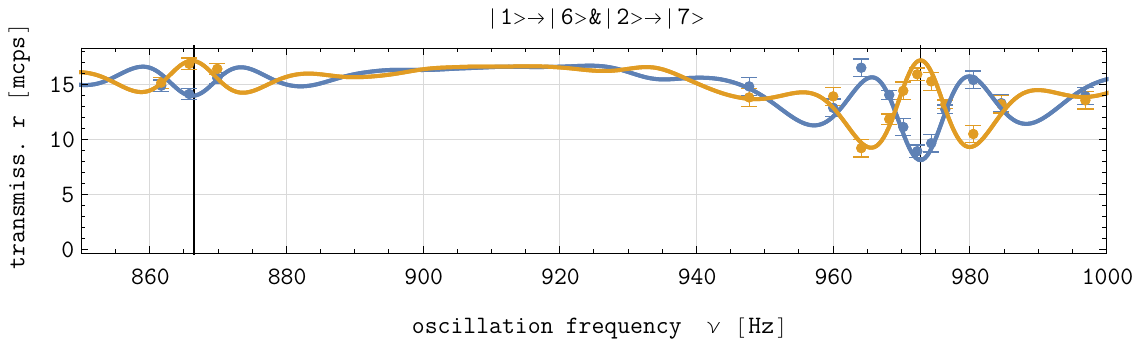}
\caption{Frequency sweep at the optimised oscillation strength. The transition frequencies of \ibc{1}{6} and \ibc{2}{7} are indicated by the black lines. The solid lines show the theory function fitted to both transitions.}\label{fig:combinedfsweep}
\end{figure}

Then the vibration strength was fixed and the frequency was varied. Again both $0$ and $\pi$ phase measurements were taken. The $\pi$ phase provides an inverted transition probability as can be seen in orange in Fig. \ref{fig:aw16sweep} and Fig. \ref{fig:combinedfsweep}. 

This complete dataset can then be fitted to the theory to determine the transition frequency $\nu_{16}$. The value for the local acceleration $g$ can be determined from the transition frequency. In addition the transition \ibc{2}{7} was measured. This also provides a check on possible shifts for a single transition. The theory gives $\nu_{27}=\SI{865.821}{\hertz}$ for the transition frequency. Fitting both of these results in Fig. \ref{fig:combinedfsweep} with parameters for the fit given in Tab. \ref{fig:fitresult}. This results in $g=\SI{9.812\pm0.0018}{\meter\per\second^2}$ which is $3.94\sigma$ off the classical measured value of $g_c=\SI{9.804925}{\meter\per\second^2}$. The value for $g_c$ was determined in 2021 next to the \qbounce experiment.

\begin{table}
\centering\caption{Resulting parameters for the fit to the complete dataset.}\label{fig:fitresult}
\begin{tabular}{|c|c|c|c|}
\hline Parameter&Result&Stat. error (1$\sigma$)&Units\\\hline 
P(1)&0.61&0.02&[1]\\
P(2)&0.22&0.03&[1]\\
P(3)&0.04&0.04&[1]\\
P$_{\mathrm{bg}}$&0.14&-&[1]\\
g&9.812&0.0018&m/s$^2$\\
r$_0$&17.27&0.165&mcps\\\hline 
$\chi^2$&74.4& & \\
n$_{\mathrm{eff}}$&73& & \\\hline 
\end{tabular}
\end{table}
\section{Systematic effects}
\qbounce uses high precision frequency measurements to determine the transition frequency between gravity states. To do this a Rb-clock is used as a frequency standard and an interferometer monitors the applied oscillation to the mirrors with a high accuracy. Nevertheless, there are frequency shifts observable for such measurements. One advantage of \qbounce is the fact that the energy splitting of states is dependent on the local acceleration $g$ and not a controllable quantity such as an applied B-Field as in NMR. Intrinsical shifts still present for \qbounce are the Bloch-Siegert shift, the spectator state shift, and Coriolis forces arising from the movement of the neutron beam.
\subsection{Spectator state shift}
Equation (\ref{eq:ramsey}) uses a two state system for illustrative purposes. In reality there are infinitely many states contributing to the transition probability. This leads to a shift of the theoretically predicted curve when compared to the simple two state case, called Spectator state shift and arises from neighbouring transitions that are weakly addressed by a driven transition \cite{baessler2015}. It has to be evaluated numerically and is taken into account by simulating a 30 state system. For \ibc{1}{6} this shift is $\Delta\nu\approx\SI{0.03}{\hertz}$ when compared to the simple two state analysis. 
\subsection{Bloch-Siegert shift}
In deriving (\ref{eq:ramsey}) the mirror movement $z(t)=a\sin(\omega t+\phi)$ was split into terms containing $e^{+i\omega t}$ and $e^{-i\omega t}$ terms. In combination with the Eigenfrequencies of the states there is one fast and one slow oscillating term. In the so called rotating wave (RWA) or secular approximation the fast oscillating term is dropped. This results in a frequency shift when compared to the full model \cite{shirley1963}. Evaluating this shift for the transition \ibc{1}{6} gives $\Delta\nu\approx\SI{15}{\milli\hertz}$. In the numerical 30-state simulation the excitation is not split and includes this shift.
\subsection{Rotation of the Earth}
By far the largest contribution to the acceleration is the centrifugal force at the surface of the Earth. This is the classical correction necessary going from a stationary to a rotating frame. In the case of the \qbounce experiment the neutrons are moving as well, which leads to an additional shift due to Coriolis forces. All apparent effects in the rotating frame can be calculated by transforming the Schrödinger equation to a rotating frame \cite{Anandan2004}. This results in the Hamiltonian
\begin{equation}
\hat{H}=\frac{{\bf{p}}^2}{2m_i}-{\bf{L}}\cdot{\bf{\Omega}}+V({\bf{r}})\;
\end{equation}
where ${\bf p}$ is the momentum vector of the neutron, ${\bf{\Omega}}$ is the angular velocity of the Earth and ${\bf{L}}$ is the angular momentum of the neutron. The gravity potential $V({\bf{r}})=-\frac{m_g r_E^2 g_0}{r}\approx-m_g r_E g_0+m_g g_0 (r-r_E)+\mathcal{O}\lb(r-r_E)^2\rb$ where $r_E$ is the radius of the Earth, $m_g$ the gravitational mass of the neutron and $g_0$ is the acceleration on the surface of the Earth. $-{\bf{L}}\cdot{\bf{\Omega}}$ is constant for a non accelerated neutron. Expanding the contribution from angular momentum to first order $\frac{{\bf{L}}^2}{2m_ir^2}\approx-\frac{L^2}{2m_ir_E^2}+\frac{L^2}{2m_i r_E^3}(r-r_E)+\mathcal{O}\lb(r-r_E)^2\rb$, where $L^2$ is the total angular momentum squared. Extracting the linear part and plugging in the expression for $L^2$ gives
\begin{equation}
g=g_0-\frac{L^2}{m_i m_gr_E^2}=g_0-\frac{m_i}{m_g}\lb2\omega v_y \sin(\chi)+r_E\omega^2\sin^2(\chi)+\frac{v^2}{r_E}\rb
\end{equation}
where $\omega$ is the angular velocity of the Earth, $\chi$ is the angle of the position of the particle to the north pole ($\chi=\pi-\mathrm{latitude}$), $v_y$ is the velocity of the neutron in western direction and $v^2$ is the square of the velocity in the lab frame. The velocity independent terms are included in a stationary measurement of the local gravitational acceleration, whereas the Coriolis force applies to the neutrons since they are moving. For \qbounce at the ILL in Grenoble we have $\chi\approx\frac{\pi}{2}$, $v_y\approx-\sin(0.33)\;\SI{8}{\meter\per\second}\approx\SI{2.6}{\meter\per\second}$, $\omega\approx\frac{2\pi}{86400}\approx\SI{7.3e-5}{\per\second}$ and $r_E\approx\SI{6.37e6}{\meter}$. Now for $m_i=m_g$ this gives
\begin{equation}
g\approx g_0-\lb-\SI{2.6e-4}{}+\SI{1.68e-2}{}+\SI{1e-5}{\meter\per\second^2}\rb\approx g_0-\SI{1.66e-2}{\meter\per\second^2}\;.
\end{equation}
For UCN with $v=\SI{8}{\meter\per\second}$ in the lab frame, the additional contributions from the Coriolis and centrifugal force attributed to the proper motion is suppressed by two orders of magnitude. 
\subsection{Phase offset}
From equation (\ref{eq:ramseyprob}) an uncertainty $\delta\phi$ in the relative phase $\Delta\phi$ between section II and IV leads directly to a shift in the transition frequency
\begin{equation}
\delta\nu_0= -\frac{\delta\phi}{2\pi\lb T+\frac{\tan\lb\frac{a\omega V_{pq}}{2}\tau\rb}{\frac{a\omega V_{pq}}{2}}\rb}\,.
\end{equation}
For a perfectly tuned $\pi/2$-flip, $a\omega V_{pq}\tau=\frac{\pi}{2}$. The average neutron has $v\approx\SI{8}{\meter\per\second}$, $T=\frac{0.34}{v}\approx\SI{0.0425}{\second}$ and $\tau=\frac{0.152}{v}\approx\SI{0.019}{\second}$. The relative phase is measured with a decoupled interferometer which reaches an accuracy of $|\delta\phi|\leq\SI{1}{\degree}$ which results in
\begin{equation}
\delta\nu_0\leq-\frac{\delta\phi}{0.419}\approx\SI{\pm4.16e-2}{\hertz}
\end{equation}
The gravitational acceleration has been measured by a stationary measurement with a falling corner cube. This is the effective acceleration also including the rotation of the Earth. From this the movement by the neutron and the quantum nature of the measurement introduce additional shifts. A summary of the described systematic shifts can be seen in Tab. \ref{fig:systematictable} where in the last line the difference to the corner cube measurement is shown. 

\begin{table}[!h]\caption{A summary of the discussed systematic effects. In the last line the expected deviation from reference measurement using a corner cube is given. This is valid only when comparing to the two state approximation, in an analysis using a multi-state system some of the shifts are already included.}\label{fig:systematictable}
\begin{tabular}{|c|c|c|}
\hline \bf Effect& $\bf\delta g$ [\si{\meter\per\second^2}]&$\bf\delta f$ [Hz]\\\hline 
Rotation of the Earth (centrifugal)&$-1.6\cdot10^{-2}$&$-1.06$\\
Phase offset&$\pm6.2\cdot10^{-4}$&$\pm4.16\cdot10^{-2}$\\
Spectator shift&$4.5\cdot10^{-4}$&$3\cdot10^{-2}$\\
Coriolis (movement of the neutron)&$2.6\cdot10^{-4}$&$1.7\cdot10^{-2}$\\
Bloch-Siegert shift&$2.3\cdot10^{-4}$&$1.5\cdot10^{-2}$\\
Centrifugal (movement of the neutron)&$-1\cdot10^{-5}$&$-6.6\cdot10^{-4}$\\\hline 
{\bf Total}&$-1.5\cdot10^{-2}\pm6.2\cdot10^{-4}$&$-1\pm4.16\cdot10^{-2}$\\
{\bf Total} ($\Delta$ to corner cube measurement)&$9.3\cdot10^{-4}\pm6.2\cdot10^{-4}$&$6.1\cdot10^{-2}\pm4.16\cdot10^{-2}$\\\hline 
\end{tabular}
\end{table}

\section{Conclusion}
The \qbounce experiment has used Ramsey GRS to probe the gravity states of neutrons bound to the surface of mirrors using mechanical oscillations to induce transitions. The transition \ibc{1}{6} was observed in detail together with \ibc{2}{7}. A transition frequency of $\nu_{16}=\SI{972.81\pm0.12}{\hertz}$ was determined which corresponds to $g=\SI{9.812\pm0.0018}{\meter\per\second^2}$. A selection of systematic effects contributing to the transition frequency was described and their magnitude is given in Fig. \ref{fig:systematictable}. The value found for the local acceleration determined by \qbounce for this dataset as compared to a classical measurement of $g_c=\SI{9.8049}{\m\per\second^2}$ using a falling corner cube deviates by a systematic offset of $\delta g\approx3.94\sigma$.

\section*{Acknowledgments}
We acknowledge support from the Austrian Fonds zur Förderung der Wissenschaftlichen Forschung (FWF-P33279). In addition we would like to thank the technical support of Thomas Brenner from the ILL staff for their invaluable assistance during the measurements. The data is available following the ILL datapolicy at https://doi.ill.fr/10.5291/ILL-DATA.3-14-412.

\end{document}